\documentclass[a4paper, 10pt, journal]{ieeeconf}  

\usepackage{amsmath}
\usepackage{amsfonts}
\usepackage{float}
\usepackage{algorithm}
\usepackage{algpseudocode}
\usepackage{booktabs}
\usepackage{siunitx}  
\usepackage{multirow}
\usepackage{dashrule}
\usepackage{graphicx}

\title{\LARGE \bf Reinforcement Learning for Null-Space Exploitation in Pseudo-Inverse Control Allocation}

\author{Umut Demir$^{1}$,  Torbj\o rn Cunis$^{1}$, Aamir Ahmad$^{1}$ and Walter Fichter$^{1}$
\thanks{$^{1}$University of Stuttgart, Faculty of Aerospace Engineering and Geodesy, Institute of Flight Mechanics and Control (iFR), Germany.
        {\tt\small firstname.lastname@ifr.uni-stuttgart.de}}}

\begin{document}

\maketitle
\thispagestyle{empty}
\pagestyle{empty}

\begin{abstract}
Pseudo-inverse methods are widely used to solve control allocation problems for overactuated aircraft with redundant control effectors. These methods reliably satisfy the primary command tracking objective, while the remaining null-space degrees of freedom are in principle available for secondary objectives such as drag reduction or actuator wear balancing. However, exploiting this freedom in practice requires hand-tuned weighting matrices or a priori preferred deflections that are difficult to specify across a wide flight envelope with nonlinear, state-dependent dynamics. This paper presents a hybrid architecture in which a reinforcement learning agent operates exclusively within the null space of the control effectiveness matrix, learning secondary objective strategies directly from interaction with the flight dynamics. The null-space projector guarantees by construction that primary command tracking remains invariant regardless of the agent output, providing a principled separation between certified primary control and adaptive secondary optimization. The approach is demonstrated on a nonlinear six-degree-of-freedom distributed electric propulsion aircraft model with aero-propulsive effects across diverse mission scenarios, achieving measurable drag reduction compared to the baseline allocation while preserving tracking integrity and respecting actuator constraints. The proposed architecture forms a foundation for future extensions toward multi-objective, mission-conditioned policies.
\end{abstract}

\section{INTRODUCTION}\label{sec:introduction}
\begin{figure}[ht]
    \centering
    \includegraphics[width=0.99\linewidth]{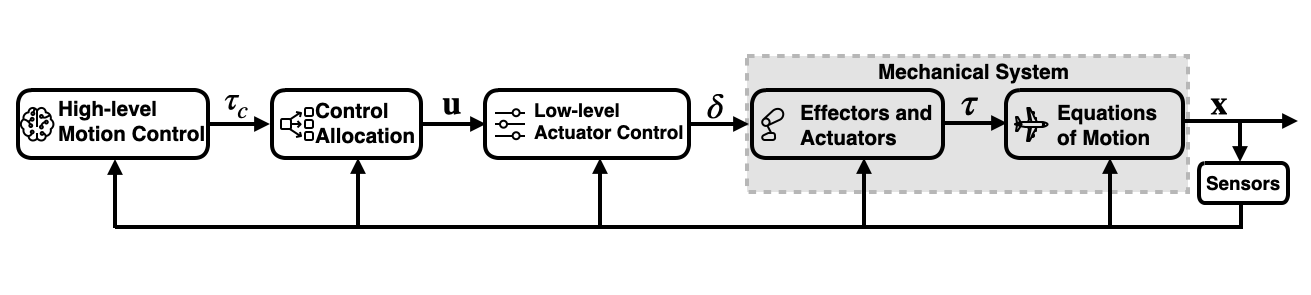}
    \caption{Control system architecture with control allocation. The high level 
    motion controller commands a virtual effort vector $\tau_c$; the allocation 
    layer distributes this demand across individual effectors}\label{fig:highlevelCA}
\end{figure}

Modern engineered systems are increasingly designed with more control effectors than independent degrees of freedom to be controlled, resulting in overactuated configurations. Examples range from autonomous robots with several effectors to satellites equipped with clusters of reaction wheels and fighter aircraft with redundant aerodynamic surfaces~\cite{yin2016review,4124854}. This redundancy is purposeful by design: a larger actuator set improves fault tolerance and reconfiguration capability, and is often preferred for reasons of safety, accuracy, bandwidth, and maintainability. In complex platforms, effector groups may serve multiple control objectives simultaneously; for example, in aircraft, engines are primarily designed for thrust yet can be commanded to generate a yawing moment when conventional aerodynamic surfaces are insufficient or unavailable~\cite{durham2017aircraft,CAsurvey}. Control allocation constitutes the intermediate layer between the high-level controller and the individual actuators, translating a commanded virtual input such as a desired moment vector into a feasible set of effector deflections while respecting physical position and rate constraints as illustrated in Fig.~\ref{fig:highlevelCA}

When the actuator set is redundant, the allocation problem admits infinitely many solutions, and the surplus degrees of freedom can in principle be directed toward secondary objectives such as minimizing drag, reducing power consumption, limiting actuator wear, or preserving a low radar cross-section. Classical methods including pseudoinverse-based allocation~\cite{durham2017aircraft}, null-space augmentation, and iterative redistribution strategies~\cite{daisy} reliably solve the primary task, but secondary objective optimization remains dependent on fixed weighting schemes or manual tuning that do not generalize across flight conditions and mission requirements. When multiple competing objectives must be balanced under nonlinear, state dependent dynamics, a learning based approach that discovers effective allocation strategies directly from interaction with the flight dynamics becomes attractive.

\subsection{Related Work}\label{subsec:related_work}

The decomposition of the control allocation solution into a pseudo-inverse particular solution and a null-space component was first formalized by Bordignon and Durham~\cite{bordignon1995null,durham1993constrained}, who introduced the null-space intersection method for constrained allocation of overactuated aircraft. Durham et al.~\cite{durham1997minimum} subsequently exploited this freedom for drag minimization by projecting cost gradients into the null space of the control effectiveness matrix. The comprehensive treatment by Durham, Bordignon, and Beck~\cite{durham2017aircraft} established the geometric theory relating the null-space structure to the Attainable Moment Set.

Several authors have applied null-space methods to specific secondary objectives in flight control. Stolk and de Visser~\cite{stolk2019minimum} developed minimum-drag control allocation for the Innovative Control Effector (ICE) tailless fighter, combining weighted pseudo-inverse prioritization with gradient descent in the null space toward minimum-drag deflections. Pfeifle and Fichter~\cite{pfeifle2023minimum} derived a closed-form null-space increment for minimum propeller shaft power in over-actuated transition aircraft. Pollack and van Kampen~\cite{pollack2023multi} extended incremental control allocation for secondary objectives using the null-space degrees of freedom.

Data-driven methods have been increasingly applied to control allocation, with neural networks trained to approximate or replace classical allocators. Kang et al.~\cite{kang2022development} train a neural allocator for a tailless aircraft that replicates a dynamic allocation scheme while avoiding heavy online optimization. Madruga et al.~\cite{madruga2021aerodynamic} use offline-trained networks to substitute the fixed allocation matrix on a quadrotor, compensating for aerodynamic effects without explicit inflow models. Chen et al.~\cite{chen2015constrained} combine an adaptive neural attitude controller with a constrained allocation layer to handle uncertainties and time-varying disturbances. Reinforcement learning extends this direction by learning allocation policies directly from interaction with the environment: de Visser et al.~\cite{de2019reinforcement} apply RL to distribute control commands across 13 effectors for longitudinal control without a preidentified effectiveness matrix. However, these approaches are fundamentally black-box function approximators in which the mapping from state to actuator command is not transparent. Secondary objectives and constraints are enforced indirectly through reward or loss shaping, behavior under distribution shift is difficult to bound, and failure diagnosis is hindered by limited interpretability. In safety-critical applications such as flight control, these limitations present significant barriers to verification and certification that often outweigh empirical performance gains.

\begin{figure*}[ht]
    \centering
    \includegraphics[width=0.75\textwidth]{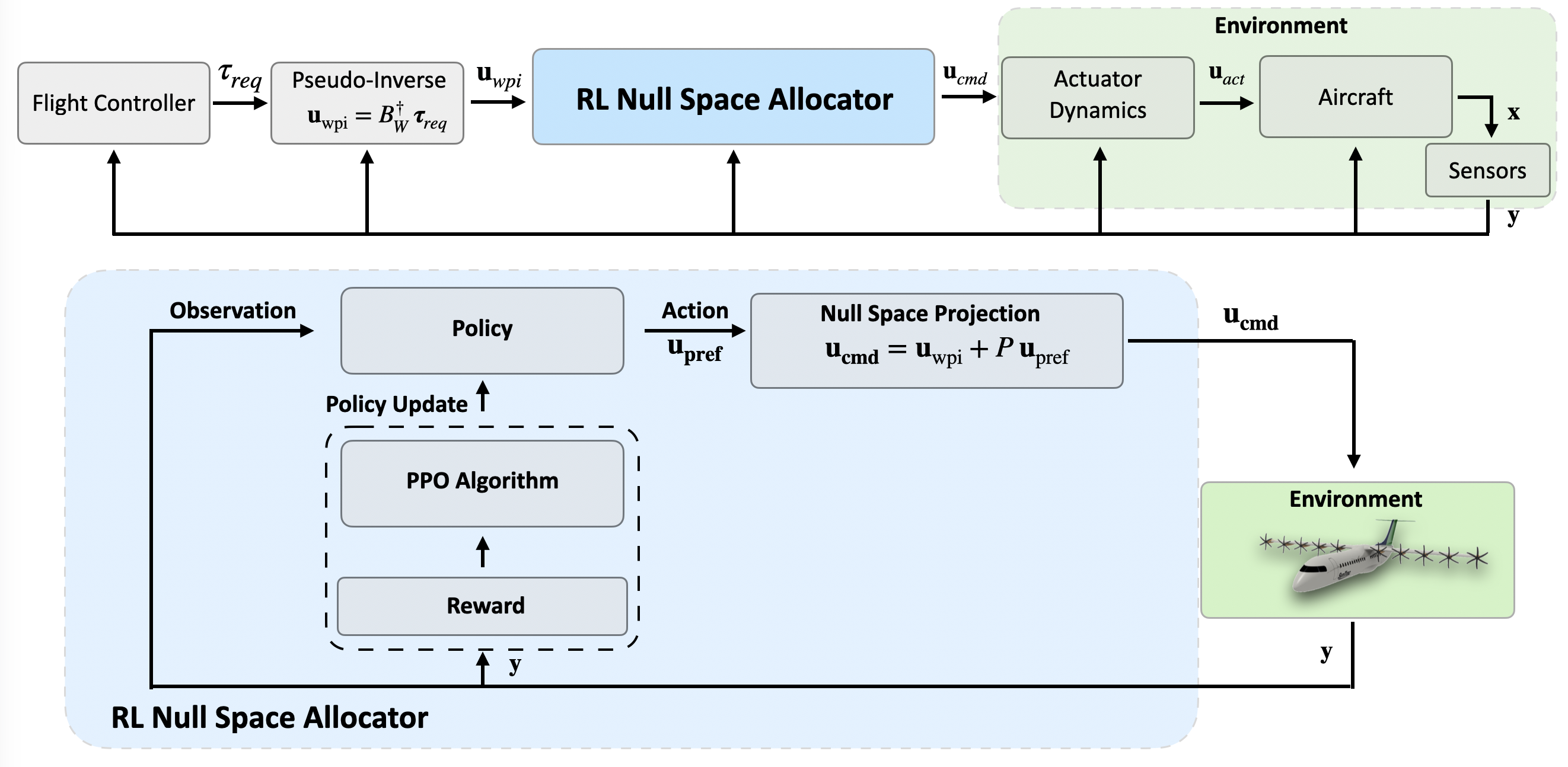}
    \caption{Overview of the hybrid control allocation architecture. The upper chain represents the conventional control loop: the flight controller generates incremental moment commands, which the Weighted Pseudo-Inverse allocator maps to a baseline actuator command. The null-space injection block modifies this baseline by adding a correction that lies entirely within the null space of the control effectiveness matrix, producing the final actuator command sent to the actuator dynamics.}
    \label{fig:architecture_overview}
\end{figure*}

While the present work was in preparation, Isci and Koyuncu~\cite{isci2026safe} recently proposed a hybrid allocation scheme for the ICE aircraft in which a reinforcement-learning agent adjusts the effector weights of an incremental quadratic programming allocator. Since a weight change only redistributes effort among solutions that satisfy the demanded moment, primary tracking is preserved while the actuator constraints remain inactive, and the agent is trained to reduce allocation error. The present work differs in the nature of the secondary objective. In their scheme the objective is internal to the allocator, the allocation error itself, whereas the secondary objectives considered here are external and mission-dependent, such as energy, actuator wear, and acoustic noise, which are nonlinear functions of the operating point rather than quantities the allocator already represents. The two directions are therefore complementary: their weight adaptation refines the allocator's internal behavior, while the present approach exploits the null space to optimize external, mission-level objectives.

This paper proposes a hybrid architecture that preserves the reliability of classical pseudo-inverse allocation for primary command tracking while employing a reinforcement learning agent to optimize the null-space component for secondary objectives. The null-space projector guarantees by construction that the learned policy cannot degrade the primary control response, regardless of the agent output. This architectural separation provides a principled safety boundary between certified primary control and adaptive secondary optimization, addressing the interpretability concerns of purely data-driven allocators while retaining the capacity to learn complex, state-dependent strategies. The approach is demonstrated on a nonlinear six-degree-of-freedom model of a distributed-electric-propulsion transport aircraft including aeropropulsive effects in a high-fidelity simulation environment.

\section{Problem Formulation}\label{sec:problem_formulation}
Consider an overactuated aircraft with $m$ control effectors and $n$ controlled axes, where $m > n$. The flight control system generates a required command $\boldsymbol{\tau}_{\mathrm{req}} \in \mathbb{R}^n$ based on the flight computer demands~\cite{durham2017aircraft}. In the linear control allocation case, the relationship between the control effector deflections $\mathbf{u} \in \mathbb{R}^m$ and the achieved command is expressed as
\begin{equation}
    \boldsymbol{\tau}_c = \mathbf{B} \, \mathbf{u},
    \label{eq:moment_mapping}
\end{equation}
where $\mathbf{B} \in \mathbb{R}^{n \times m}$ is the control effectiveness matrix, whose entries $B_{ij} = \partial \tau_{c_i} / \partial u_j$ represent the partial derivatives of the achieved command with respect to each control effector deflection. The primary objective of control allocation is to find $\mathbf{u}$ such that
\begin{align}
    \boldsymbol{\tau}_c = &\;\mathbf{B} \, \mathbf{u} \label{eq:ca_primary} 
\end{align}

Since $m > n$, the system~\eqref{eq:ca_primary} is underdetermined and admits infinitely many solutions. A common approach is the Weighted Pseudo-Inverse (WPI), which yields the minimum-norm solution
\begin{equation}
    \mathbf{u}_{\mathrm{wpi}} = W^{-1} B^\top \bigl( B \, W^{-1} B^\top \bigr)^{-1} \boldsymbol{\tau}_c = B_W^\dagger \, \boldsymbol{\tau}_c,
    \label{eq:wpi}
\end{equation}
where $W \in \mathbb{R}^{m \times m}$ is a positive-definite weighting matrix and $B_W^\dagger$ denotes the weighted pseudo-inverse.

The weighted pseudo-inverse commits to a single minimum-norm allocation and thereby leaves the redundancy of the overactuated system unused. This surplus freedom becomes explicit when the solution set is expressed through the null space of $\mathbf{B}$. 
Any solution of~\eqref{eq:ca_primary} can be written as the sum of a particular solution and a homogeneous component lying in the null space of $B$~\cite{durham2017aircraft},
\begin{equation}
    \mathbf{u} = \mathbf{u}_{\mathrm{wpi}} + \mathbf{u}_{\perp},
    \label{eq:decomposition}
\end{equation}
where $\mathbf{u}_{\mathrm{wpi}} = B_W^\dagger \boldsymbol{\tau}_{\mathrm{req}}$ is the particular solution from~\eqref{eq:wpi}, and the homogeneous component satisfies $\mathbf{u}_{\perp} \in \mathcal{N}(B)$, i.e.\ $B\,\mathbf{u}_{\perp} = \mathbf{0}$, so it does not change the achieved command $\boldsymbol{\tau}_c = B\mathbf{u}$. When $B$ has full row rank $n$, the null space $\mathcal{N}(B)$ has dimension $m - n$, corresponding to the degrees of freedom that remain once the primary objective is met. This component is parameterized by projecting an arbitrary preferred deflection $\mathbf{u}_{\mathrm{pref}} \in \mathbb{R}^m$ onto $\mathcal{N}(B)$,
\begin{equation}
    \mathbf{u}_{\perp} = P \, \mathbf{u}_{\mathrm{pref}}, \qquad P = I_m - B_W^\dagger \, B,
    \label{eq:null_space_projection}
\end{equation}
where $P \in \mathbb{R}^{m \times m}$ is the null-space projector. Since $B_W^\dagger$ is a right inverse of $B$, the projector satisfies $B P = \mathbf{0}$ by construction, so the primary command-tracking objective~\eqref{eq:ca_primary} remains invariant for any choice of $\mathbf{u}_{\mathrm{pref}}$. The complete allocation law therefore becomes
\begin{equation}
    \mathbf{u} = B_W^\dagger \, \boldsymbol{\tau}_{\mathrm{req}} + P \, \mathbf{u}_{\mathrm{pref}},
    \label{eq:full_allocation}
\end{equation}
where the first term produces the required command and the second provides the $m - n$ free parameters that can be exploited for secondary objectives without affecting the primary control task.

In~\eqref{eq:full_allocation} the preferred deflection $\mathbf{u}_{\mathrm{pref}}$ is the only free quantity: it selects one point in the $(m-n)$-dimensional null space and thereby fixes the secondary behavior, while the achieved command stays invariant. Classical schemes set it a~priori through fixed weights or scheduled targets, which do not generalize across the flight envelope and cannot capture secondary objectives such as energy that are nonlinear and accumulate over the mission. We therefore let a learned policy $\pi_{\theta}$ generate the preference from the current state,
\begin{equation}
    \mathbf{u}_{\mathrm{pref}} = \pi_{\theta}(\mathbf{s}),
    \label{eq:policy}
\end{equation}
where $\mathbf{s}$ encodes the flight condition and mission, and the parameters $\theta$ are trained by reinforcement learning to maximize a cumulative return rather than an instantaneous cost. The following section formulates this learning problem.

\section{Reinforcement Learning}\label{sec:rl_agent}
Reinforcement learning (RL) formulates sequential decision-making as a Markov Decision Process (MDP), defined by the tuple $(\mathcal{S}, \mathcal{A}, \mathcal{P}, r, \gamma)$, where $\mathcal{S}$ is the state space, $\mathcal{A}$ the action space, $\mathcal{P}: \mathcal{S} \times \mathcal{A} \times \mathcal{S} \to [0,1]$ the transition probability, $r: \mathcal{S} \times \mathcal{A} \to \mathbb{R}$ the reward function, and $\gamma \in [0,1)$ the discount factor~\cite{sutton1998reinforcement}. An agent interacts with the environment by observing a state $\mathbf{s}_t \in \mathcal{S}$, selecting an action $\mathbf{a}_t \in \mathcal{A}$ according to a policy $\pi_\theta(\mathbf{a}_t \mid \mathbf{s}_t)$, and receiving a scalar reward $r_t$. The objective is to find the parameters $\theta$ that maximize the expected discounted return
\begin{equation}
    R(\theta) = \mathbb{E}_{\pi_\theta} \left[ \sum_{t=0}^{T} \gamma^t \, r_t \right]
    \label{eq:rl_objective}
\end{equation}

Policy gradient methods optimize~\eqref{eq:rl_objective} by estimating the gradient of the expected return with respect to the policy parameters. Among these, Proximal Policy Optimization (PPO) has emerged as a robust and sample-efficient algorithm for continuous control tasks~\cite{schulman2017proximal}. PPO constrains the policy update to a trust region by clipping the probability ratio between the new and old policies,
\begin{equation}
    L^{\mathrm{CLIP}}(\theta) = \mathbb{E}_t \!\left[ \min\!\bigl( \rho_t \hat{A}_t,\, \mathrm{clip}(\rho_t, 1\!-\!\epsilon, 1\!+\!\epsilon)\, \hat{A}_t \bigr) \right],
    \label{eq:ppo_objective}
\end{equation}
where $\rho_t(\theta) = \pi_\theta(\mathbf{a}_t \mid \mathbf{s}_t) / \pi_{\theta_{\mathrm{old}}}(\mathbf{a}_t \mid \mathbf{s}_t)$ is the importance sampling ratio, $\hat{A}_t$ the estimated advantage function, and $\epsilon$ the clipping parameter that controls the maximum allowed policy deviation per update.

The null-space allocation problem introduced in Section~\ref{sec:problem_formulation} maps naturally onto this MDP framework. The state is an observation of the information available to the allocator at time $t$,
\begin{equation}
    \mathbf{s}_t = \mathbf{o}\!\left( \mathbf{y}_t,\ \boldsymbol{\tau}_{\mathrm{req},t},\ \mathbf{u}_t \right) \in \mathcal{S} \subset \mathbb{R}^{d_s},
    \label{eq:state_vector_rl}
\end{equation}
where $\mathbf{y}_t$ are the measured flight variables, $\boldsymbol{\tau}_{\mathrm{req},t}$ the required command, and $\mathbf{u}_t$ the current effector configuration. The observation map $\mathbf{o}(\cdot)$ selects and normalizes these signals, together with quantities derived from them such as the baseline solution $\mathbf{u}_{\mathrm{wpi},t}$ or the null-space coordinates. The specific observation used in this work is given in Section~\ref{subsec:aircraft_model}. The action is the preferred deflection from~\eqref{eq:policy},
\begin{equation}
    \mathbf{a}_t = \mathbf{u}_{\mathrm{pref},t} = \pi_{\theta}(\mathbf{s}_t) \in \mathcal{A} \subset \mathbb{R}^m,
    \label{eq:action_rl}
\end{equation}
which the null-space projector maps to the moment-invariant correction $P\,\mathbf{u}_{\mathrm{pref},t}$ before it is added to the baseline solution \eqref{eq:full_allocation}.
which the null-space projector maps to the moment-invariant correction $P \, \mathbf{u}_{\mathrm{RL},t}$ before it is added to the baseline solution as in~\eqref{eq:full_cmd}. The projection enforces $B \, P \, \mathbf{u}_{\mathrm{RL},t} = \mathbf{0}$ regardless of the agent output, effectively constraining the action space to $\mathcal{N}(B)$ without requiring explicit enforcement during training. The state transition
\begin{equation}
    \mathbf{s}_{t+1} \sim \mathcal{P}(\cdot \mid \mathbf{s}_t, \mathbf{a}_t)
    \label{eq:transition_rl}
\end{equation}
is governed by the nonlinear six-degree-of-freedom aircraft dynamics $\mathbf{f}(\mathbf{x}, \mathbf{u})$, the actuator models, and the flight controller response. A mission-conditioned reward
\begin{equation}
    r_t = r(\mathbf{s}_t, \mathcal{M}), \qquad \mathcal{M} \in \{\text{energy}, \text{actuator wear}, \text{noise}\},
    \label{eq:reward_rl}
\end{equation}
evaluates the resulting allocation with respect to the active mission profile $\mathcal{M}$ and provides the scalar signal for the PPO policy update~\eqref{eq:ppo_objective}. The agent thereby learns to select the preferred deflection $\mathbf{u}_{\mathrm{pref}} = \pi_\theta(\mathbf{s})$ so that the secondary objective is optimized over the course of a mission.

\section{Training Setup}\label{sec:training}

\subsection{Aircraft Model}\label{subsec:aircraft_model}

Having established the general control allocation framework for an arbitrary overactuated system with $m$ effectors and $n$ controlled axes, the formulation is now applied to a specific aircraft configuration. The SynTrac Configuration~2a distributed electric propulsion (DEP) aircraft serves as the application platform~\cite{schlittenhardt2025preliminary}. It is a commuter-class transport whose overactuation arises from the combination of conventional aerodynamic control surfaces and ten wing-mounted propellers, giving a large actuator redundancy relative to the three controlled moment axes. The rigid-body dynamics are modeled as a six-degree-of-freedom (6-DOF) nonlinear system with state vector

\begin{figure}[htbp]
    \centering
    \includegraphics[width=0.85\columnwidth]{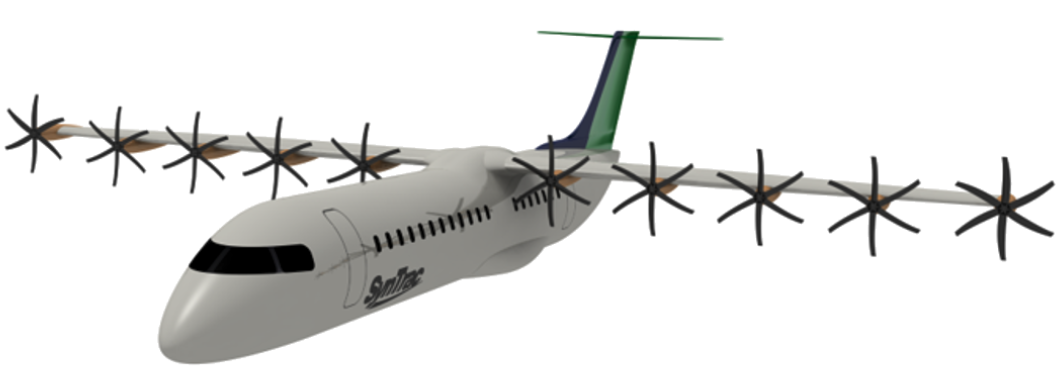}
    \caption{The SynTrac Configuration~2a distributed electric propulsion aircraft. The control effectors comprise four aerodynamic surfaces (left and right ailerons, elevator, and rudder) and the blade pitch angles of ten wing-mounted propellers.}
    \label{fig:syntrac_config2a}
\end{figure}
\begin{equation}\label{eq:state_vector}
    \mathbf{x} = \bigl[\, V \;\; \alpha \;\; \beta \;\; p \;\; q \;\; r \;\; \phi \;\; \theta \;\; \psi \;\; x_E \;\; y_E \;\; z_E \,\bigr]^\top,
\end{equation}
where $V$ is the true airspeed, $\alpha$ the angle of attack, $\beta$ the sideslip angle, $p$, $q$, $r$ the body-axis roll, pitch, and yaw rates, $\phi$, $\theta$, $\psi$ the Euler angles, and $x_E$, $y_E$, $z_E$ the position coordinates in the Earth-fixed frame. The principal geometric and inertial properties of the Configuration~2a aircraft are listed in Table~\ref{tab:syntrac_params}.

\begin{table}[htbp]
    \centering
    \caption{Configuration parameters of the SynTrac Configuration 2a aircraft (MTOM case).}
    \label{tab:syntrac_params}
    \begin{tabular}{l c c}
        \toprule
        Parameter & Value & Unit \\
        \midrule
        Wing area, $S$          & $58.4$   & m$^2$ \\
        Wing span, $b$          & $27.00$  & m \\
        Mean wing chord, $\bar{c}$ & $2.16$ & m \\
        Mass, $m$               & $24533$  & kg \\
        \midrule
        $I_x$                   & $217205$ & kg\,m$^2$ \\
        $I_y$                   & $612834$ & kg\,m$^2$ \\
        $I_z$                   & $705546$ & kg\,m$^2$ \\
        $I_{xz}$                & $0$      & kg\,m$^2$ \\
        \bottomrule
    \end{tabular}
\end{table}

The aircraft is controlled through fourteen effectors: four aerodynamic control surfaces, namely a left and right aileron ($\delta_{a_L}$, $\delta_{a_R}$), an elevator ($\delta_{e}$), and a rudder ($\delta_{r}$), together with the blade pitch angles of ten wing-mounted, constant-speed propellers ($\delta_{p_1}, \dots, \delta_{p_{10}}$), whose thrust is modulated through pitch. These are collected in the effector command vector
\begin{equation}\label{eq:actuator_vector}
    \mathbf{u} = \bigl[\, \delta_{a_L} \;\; \delta_{a_R} \;\; \delta_{e} \;\; \delta_{r} \;\; \delta_{p_1} \;\; \cdots \;\; \delta_{p_{10}} \,\bigr]^\top \in \mathbb{R}^{14},
\end{equation}
subject to the position and rate constraints summarized in Table~\ref{tab:actuator_limits}. Each actuator is modeled as a second-order system with natural frequency $\omega_i$ and damping ratio $\zeta_i = 1$ (critically damped).

\begin{table}[htbp]
    \centering
    \caption{SynTrac Configuration 2a effector constraints and dynamics.}
    \label{tab:actuator_limits}
    \begin{tabular}{l c c c c}
        \toprule
        & $u_{\min}$ [$^\circ$] & $u_{\max}$ [$^\circ$] & $\dot{u}_{\max}$ [$^\circ$/s] & $\omega$ [rad/s] \\
        \midrule
        $\delta_{a_L}$          & $-14$ & $14$ & $20$  & $30$ \\
        $\delta_{a_R}$          & $-14$ & $14$ & $20$  & $30$ \\
        $\delta_{e}$            & $-23$ & $13$ & $20$  & $30$ \\
        $\delta_{r}$            & $-25$ & $25$ & $20$  & $30$ \\
        $\delta_{p_1}\!-\!\delta_{p_{10}}$ & $-15$ & $15$ & $20$ & $10$ \\
        \bottomrule
    \end{tabular}
\end{table}

\subsection{SynTrac Flight Controller and Control Allocation}
\label{subsec:controller_ca}

The inner loop is based on incremental nonlinear dynamic inversion (INDI). An outer attitude and rate controller generates the required angular acceleration $\boldsymbol{\tau}_{\mathrm{req}} \in \mathbb{R}^{3}$ from the tracking error, and the INDI law realizes it incrementally about the current operating point,
\begin{align}
    \mathbf{u}_{\mathrm{cmd}} &= \mathbf{u}_0 + B_W^\dagger \bigl( \boldsymbol{\tau}_{\mathrm{req}} - \boldsymbol{\tau}_0 \bigr) + P \, \mathbf{u}_{\mathrm{pref}}
    \label{eq:indi_allocation}
\end{align}
where $\mathbf{u}_0$ is the current measured effector configuration and $\boldsymbol{\tau}_0$ the angular acceleration measured at that configuration. The first two terms form the weighted pseudo-inverse increment that tracks the required command, and the third injects the learned null-space preference. Equation~\eqref{eq:indi_allocation} is the INDI realization of the general allocation law~\eqref{eq:full_allocation}, in which the particular solution is taken incrementally about the current state.

The control effectiveness matrix is evaluated at the current operating point, so it captures the state dependence of both the aerodynamic surfaces and the propeller pitch, including the aeropropulsive coupling of the distributed propulsion system. Each column $B_{\cdot j} = \partial \boldsymbol{\tau}_c / \partial u_j$ is obtained from the onboard model by perturbing effector $j$ about its nominal value, which avoids precomputed gain tables and keeps the allocation valid across the flight envelope.

\begin{table}[htbp]
    \centering
    \caption{PPO agent architecture and training hyperparameters.}
    \label{tab:ppo_config}
    \begin{tabular}{l c}
        \toprule
        \textbf{Actor Network} & \\
        \midrule
        Normalization & Scaling layer \\
        Hidden layers & 64, 64, 32 \\
        Activations & ReLU, ReLU, tanh \\
        Mean output & Fully Connected $\rightarrow$ tanh $\rightarrow$ scaling \\
        Std. output & Fully Connected $\rightarrow$ softplus $\rightarrow$ scaling \\
        \midrule
        \textbf{Critic Network} & \\
        \midrule
        Normalization & Scaling layer \\
        Hidden layers & 64, 32 \\
        Activations & ReLU, ReLU \\
        Output & Fully Connected \\
        \midrule
        \textbf{PPO Hyperparameters} & \\
        \midrule
        Clip factor $\epsilon$ & 0.15 \\
        Discount factor $\gamma$ & 0.995 \\
        Mini-batch size & 64 \\
        Entropy loss weight & 0.01 \\
        Actor learning rate & $1 \times 10^{-4}$ \\
        Critic learning rate & $1 \times 10^{-5}$ \\
        Gradient threshold & 0.5  \\
        \bottomrule
    \end{tabular}
\end{table}

\subsection{Training}
\label{subsec:training}

The policy and value function are represented by the two feed-forward networks summarized in the upper part of Table~\ref{tab:ppo_config}. The actor is a continuous Gaussian policy that maps the observation $\mathbf{s}$ to the mean and standard deviation of the action $\mathbf{a} = \mathbf{u}_{\mathrm{pref}}$. It uses three hidden layers of $64$, $64$ and $32$ units with ReLU and tanh activations. A scaling layer normalizes the observation at the input; the mean head is passed through a tanh and a scaling layer so that the commanded preference stays within the admissible effector range, and the standard-deviation head uses a softplus to remain positive. The critic is a smaller network, with hidden layers of $64$ and $32$ units, that estimates the state value and serves as the baseline in the advantage computation.

Each training episode samples a mission from a library that spans the flight envelope and the operational mission set, with every mission initialized from trimmed flight. The reward is scaled to order one so that the returns stay in a numerically favorable range, and the experience horizon is matched to the episode length, which keeps the advantage estimates consistent along the trajectory. At each step the policy proposes a preferred deflection, the null-space projector maps it to the moment-invariant correction, and the resulting allocation is evaluated by the mission-conditioned reward to form the training signal.

\section{Results}\label{sec:results}
The trained PPO agent is evaluated against the baseline WPI allocation to assess the effect of the learned null-space policy on secondary objective performance. 

\subsection{Drag Minimization}
\label{subsec:results_drag}

The trained drag-minimization policy is evaluated against the weighted pseudo-inverse baseline on a representative mission that combines a lateral-directional command at $t = 10\,\mathrm{s}$ with an angle-of-attack command at $t = 25\,\mathrm{s}$. Each case is run under identical initial conditions and reference commands, once with the agent disabled (RL off, baseline allocation) and once with the agent active (RL on), so that any difference is attributable solely to the learned null-space correction.

\begin{figure}[htbp]
    \centering
    \includegraphics[width=\linewidth]{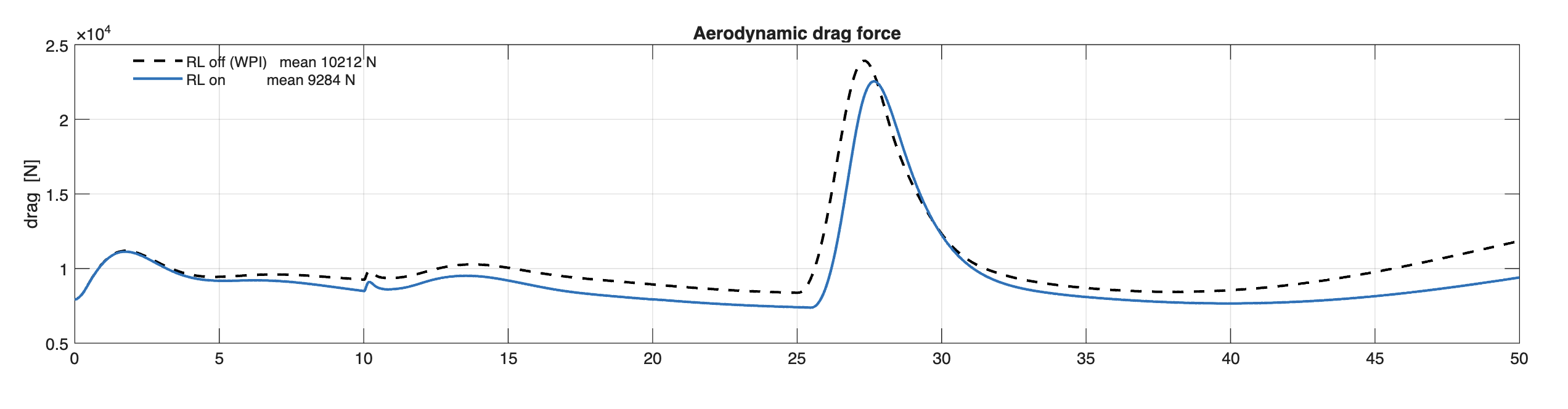}
    \caption{Aerodynamic drag force over the mission for the baseline (RL off, WPI) and the learned allocation (RL on). The mean drag is reduced from $10212\,\mathrm{N}$ to $9284\,\mathrm{N}$, a $9.1\,\%$ reduction.}
    \label{fig:drag_history}
\end{figure}

\begin{figure}[htbp]
    \centering
    \includegraphics[width=\linewidth]{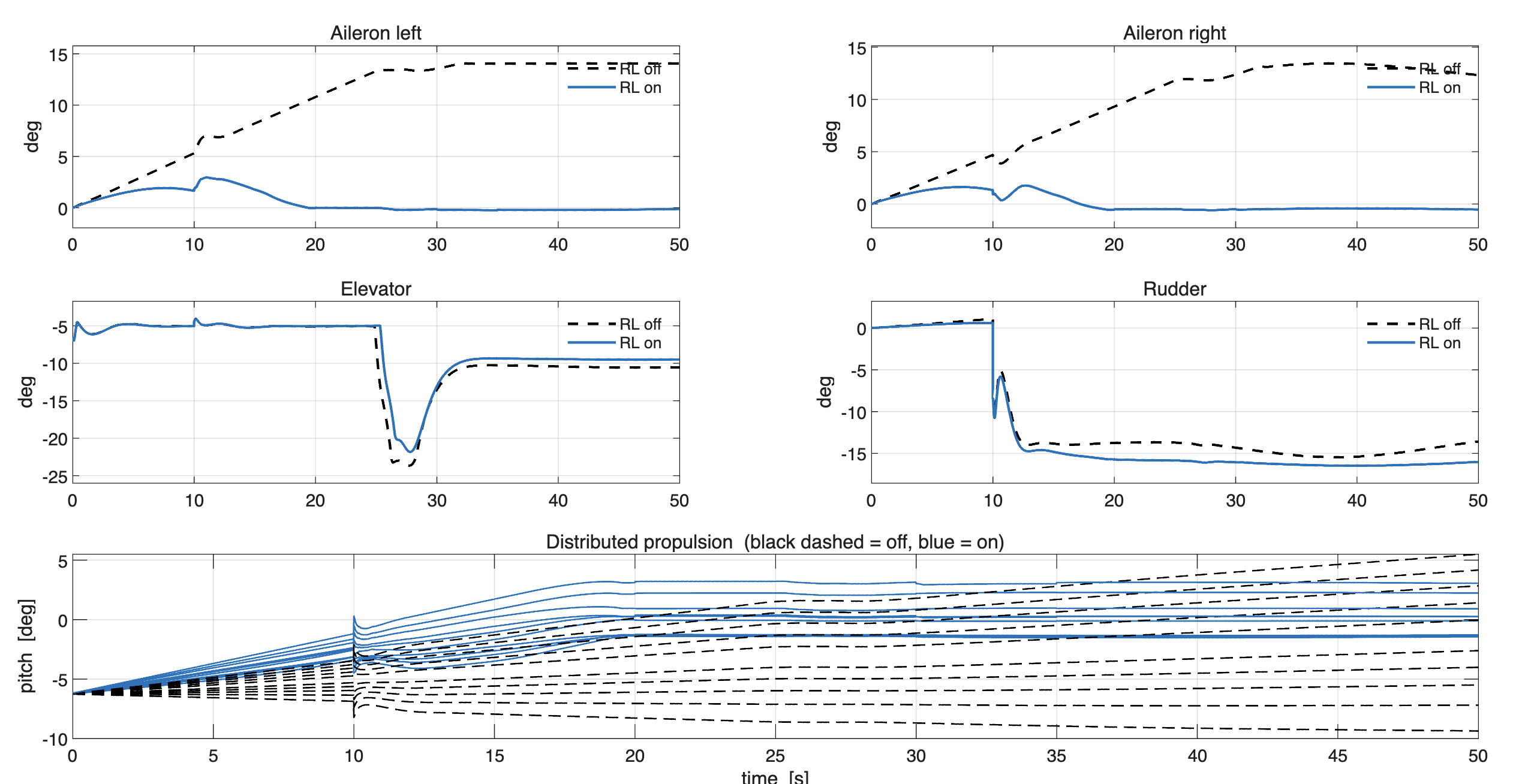}
    \caption{Effector deflections for the baseline (RL off) and the learned allocation (RL on): the four aerodynamic surfaces and the ten distributed-propulsion pitch angles. The agent holds the ailerons near zero and keeps the propeller pitch clustered, while the elevator and rudder, which carry the commanded pitch and yaw, are nearly unchanged.}
    \label{fig:drag_inputs}
\end{figure}

\begin{figure}[htbp]
    \centering
    \includegraphics[width=\linewidth]{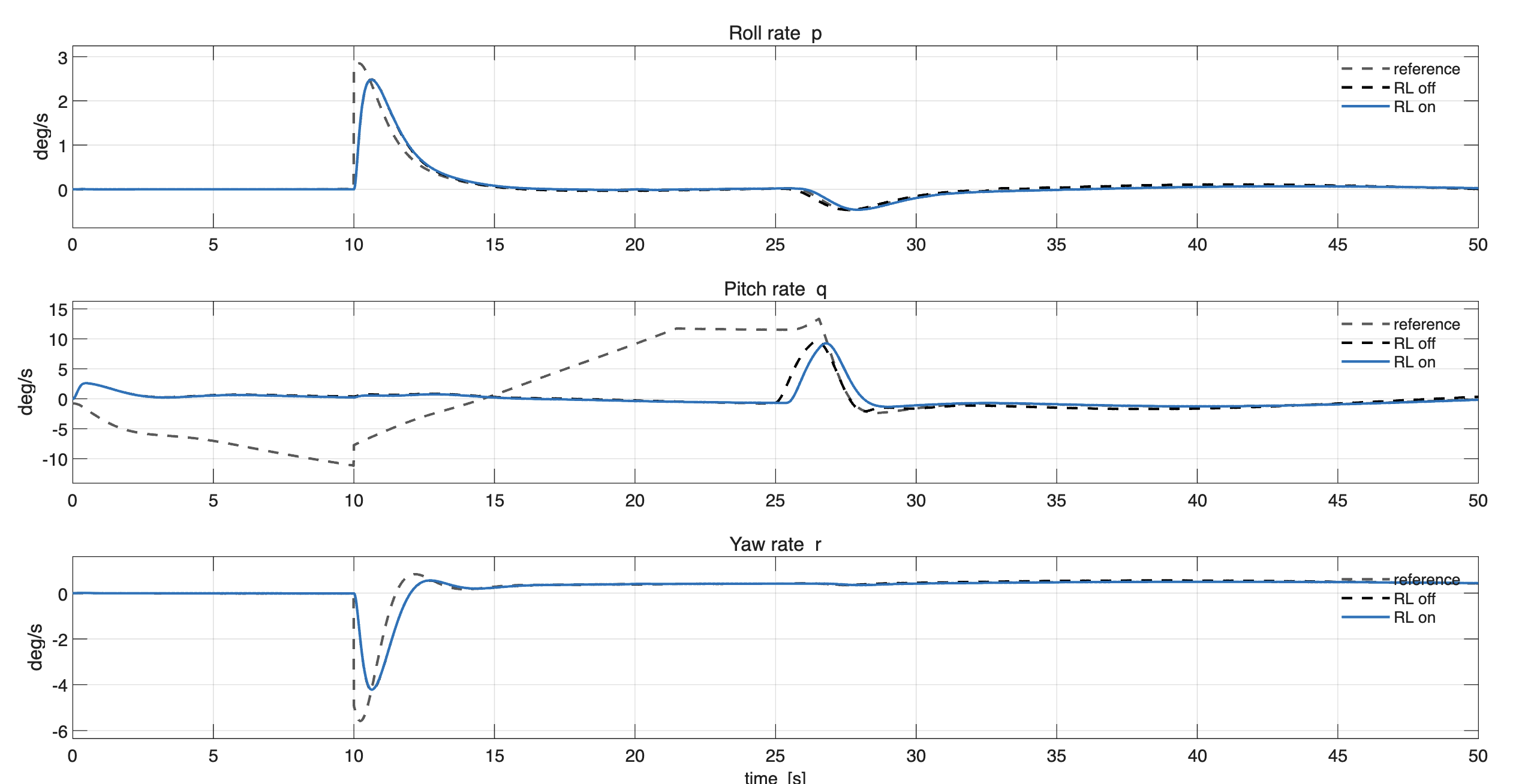}
    \caption{Body-axis roll, pitch and yaw rates for the reference, the baseline (RL off) and the learned allocation (RL on). The rates are essentially identical with and without the agent, confirming moment-tracking invariance.}
    \label{fig:drag_rates}
\end{figure}

\begin{figure}[htbp]
    \centering
    \includegraphics[width=\linewidth]{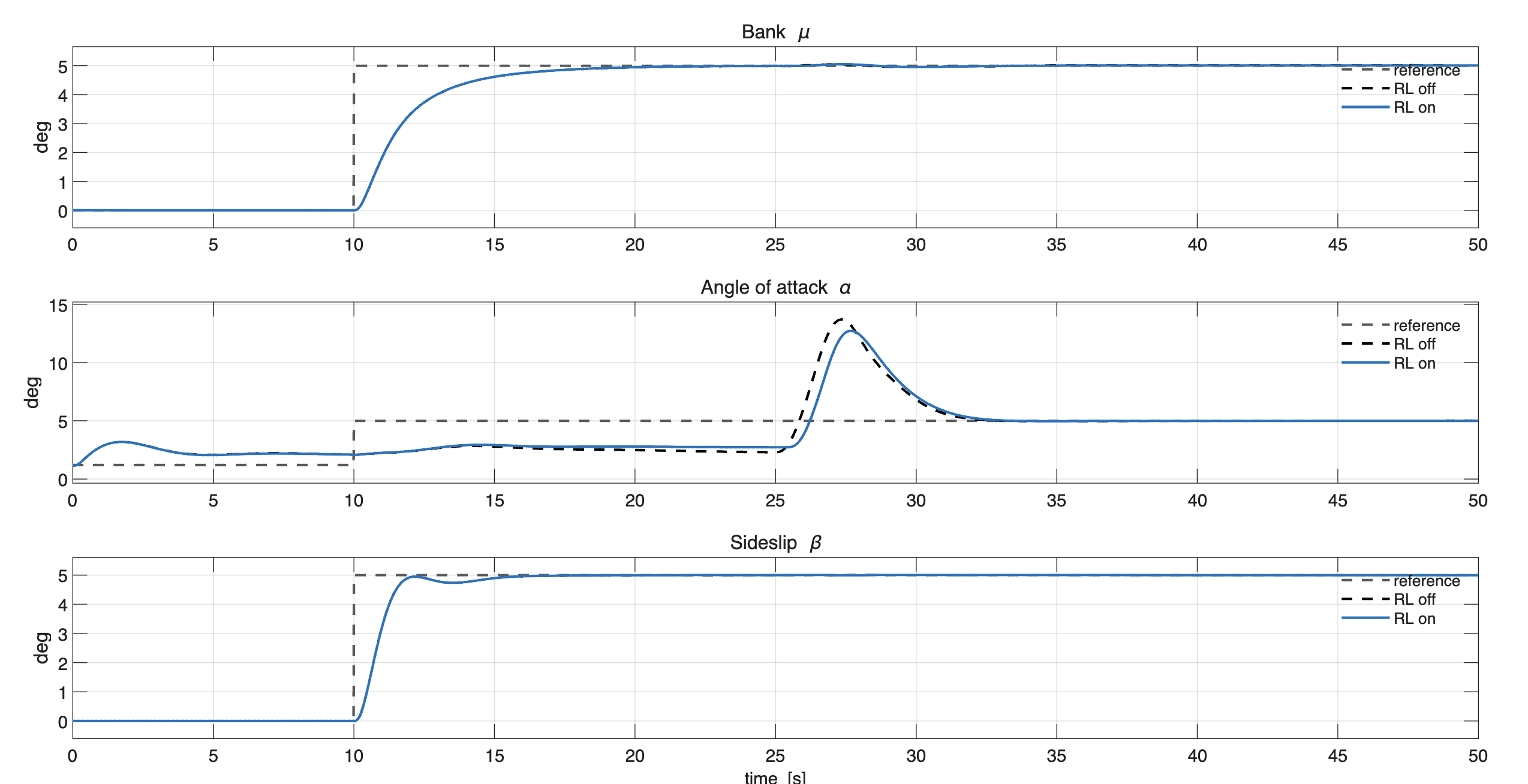}
    \caption{Bank angle $\mu$, angle of attack $\alpha$ and sideslip $\beta$ for the reference, the baseline (RL off) and the learned allocation (RL on). The attitude response is unchanged by the null-space correction.}
    \label{fig:drag_attitude}
\end{figure}

Figure~\ref{fig:drag_history} shows the aerodynamic drag force over the mission. The mean drag falls from $10212\,\mathrm{N}$ for the baseline to $9284\,\mathrm{N}$ with the agent active, a reduction of $928\,\mathrm{N}$ or $9.1\,\%$. The benefit is sustained rather than transient: the RL-on trace lies at or below the baseline throughout, with the largest relative reductions in the quasi-steady segments before $25\,\mathrm{s}$ and after $30\,\mathrm{s}$, and a lowered peak during the high-load event near $27\,\mathrm{s}$, where the drag maximum drops from about $2.4 \times 10^{4}\,\mathrm{N}$ to $2.25 \times 10^{4}\,\mathrm{N}$.

That the reduction is achieved without disturbing the primary task is confirmed by Figures~\ref{fig:drag_rates} and~\ref{fig:drag_attitude}. The roll, pitch and yaw rates (Fig.~\ref{fig:drag_rates}) and the bank, angle-of-attack and sideslip responses (Fig.~\ref{fig:drag_attitude}) are essentially indistinguishable between RL off and RL on, and both track the reference commands equally. This is the empirical counterpart of the invariance established in Section~\ref{sec:problem_formulation}: because the learned correction lies in the null space of $B$, it leaves the achieved moments, and therefore the closed-loop attitude and rate response, unchanged.

The source of the drag reduction is visible in the effector histories of Figure~\ref{fig:drag_inputs}. The baseline deploys the left and right ailerons close to their limits and spreads the ten propeller pitch angles over a wide range, whereas the learned policy holds the ailerons near zero and keeps the propeller pitch tightly clustered. The elevator and rudder, which carry the commanded pitch and yaw, are nearly identical in both cases and retain little null-space freedom. The agent therefore unloads the aerodynamic surfaces and redistributes the trim onto the distributed propulsion, reaching a less draggy configuration that produces the same moments. This is exactly the redundancy that the baseline leaves unused and that the null-space policy is designed to exploit.

\section{Conclusion and Outlook}\label{sec:conclusion}
This work presented a hybrid control allocation architecture in which a deterministic pseudo-inverse resolves the primary command while a reinforcement-learning policy acts within the null space of the control effectiveness matrix to optimize a secondary objective. Because the learned correction enters through the null-space projection, it cannot alter the achieved moments, so command tracking is preserved by construction and the certification argument is confined to the deterministic core. 

Future work will extend the approach along two directions. The first is a broader set of mission profiles, with a single mission-conditioned policy trained across the flight envelope and the operational mission set, so that the secondary objective can be adapted online and evaluated on further objectives such as acoustic noise. The second addresses the path dependency introduced by the incremental INDI allocation. Because the effectors retain their previous positions, these null-space increments accumulate from step to step. The configuration reached at any instant is therefore the sum of all past corrections rather than a function of the current command, which lets it drift toward the actuator limits and makes the achieved secondary objective depend on the path taken rather than on the instantaneous state, something an instantaneous allocation rule cannot account for.





\section*{ACKNOWLEDGMENT}
This work was funded by the Deutsche Forschungsgemeinschaft
(DFG, German Research Foundation) - Project ID 498601949 - TRR
364 SynTrac

\bibliographystyle{IEEEtran.bst}
\bibliography{IEEEexample.bib}

\end{document}